\documentclass[journal, paper]{IEEEtran}

\usepackage{amsmath}
\usepackage{amsfonts}
\usepackage{amssymb}
\usepackage{graphics}
\usepackage{empheq}
\usepackage{pstricks}
\usepackage{array}
\usepackage{float}
\usepackage{epsf}
\usepackage{graphicx}
\usepackage{multirow}
\usepackage{cite}
\usepackage{physics}
\usepackage{acronym}
\usepackage{comment}
\usepackage{upgreek}
\usepackage{subcaption}
\usepackage{nicefrac} 
\usepackage[font=small]{caption}
\usepackage{comment}
\usepackage{pifont}

\newcommand{\bfg}[1]{\boldsymbol{#1}}

\newcommand{\bfb}[1]{\boldsymbol{\rm #1}}

\newcommand{\Wt}{Wiener process}

\newcommand{\OK}{Yes}
\newcommand{\NO}{--}
\newcommand{\Dt}{\dot}

\acrodef{rocof}[RoCoF]{Rate of Change of Frequency}

\renewcommand{\arraystretch}{0.8} 

\begin{document}

\title{Asymmetry of Frequency Distribution in Power Systems: Sources, Estimation, Impact and Control}

\author{Taulant~K\"{e}r\c{c}i, \IEEEmembership{IEEE~Senior~Member,~}and Federico~Milano, \IEEEmembership{IEEE~Fellow}
  \thanks{T.~K\"{e}r\c{c}i is with EirGrid plc, Transmission System Operator, Dublin, D04 FW28, Ireland. F.~Milano is with School of Electrical and Electronic Engineering, University College Dublin, Dublin, D04V1W8, Ireland.  E-mails: taulant.kerci@eirgrid.com, federico.milano@ucd.ie.}%
}

\maketitle

\markboth{Asymmetry of Frequency Distribution in Power Systems}{T. K\'er\c{c}i and F. Milano}

\begin{abstract}
  This paper analyses an emerging real-world phenomena in inverter-based renewable-dominated power systems, namely, asymmetry of frequency distribution.  The paper first provides a rationale on why asymmetry reduces the ``quality'' of the frequency control and system operation.  Then it provides qualitative theoretical insights that explain asymmetry in terms of the nonlinearity of real-world power systems and associated models.  In particular network losses and pitch angle-based frequency control of wind power plants are discussed.  Then the paper proposes a nonlinear compensation control to reduce the asymmetry as well as a statistical metric based on the frequency probability distribution to quantify the level of asymmetry in a power system.   Real-world data obtained from the Irish and Australian transmission systems serve to support the theoretical appraisal, whereas simulations based on an IEEE benchmark system show the effectiveness of the proposed nonlinear compensation.  The case study also shows that, while automatic generation control reduces asymmetry, frequency control limits and droop-based frequency support provided by wind generation using a tight deadband of $\pm$15 mHz, namely active power control, leads to a significant increase in the asymmetry of the frequency probability distribution.
\end{abstract}


\maketitle

\section{Introduction}
\label{sec:intro}

\IEEEPARstart{I}{ntuitively}, the higher the symmetry in the dynamic response of a dynamical system, the higher the predictability of the system behavior and controllability.  Thus, a precise evaluation of the asymmetries present in a system contributes towards increased power system stability and resilience.  The equations that represent power systems are for the most part symmetrical.  However, it has been recently observed by some transmission system operators (TSOs), that power systems are becoming increasingly (and unexpectedly) asymmetrical \cite{csenergy, aemopfc}.  In this context, the aim of this paper is to study the causes of frequency probability distribution (FPD) asymmetry in power systems, provide a qualitative theoretical appraisal of this asymmetry; and, finally, show how asymmetry can be compensated through modified control.

The topic of FPD in power systems and the various sources and parameters that influence it has recently received attention in the literature, in particular, in light of the integration of uncertain and variable renewable energy sources (RES) such as wind and solar generation \cite{10015191, 9744334, DELGIUDICE2021106842, 8963682, 8626538, 7540970}.
The main focus of these works is on the modelling and study of how to reproduce the frequency distribution observed in real grids, e.g., the bi-modal distribution.  However, the effect of losses, control limits and RES providing primary frequency control (PFC) has not been considered so far.   

With regard to the latter, there is a concern in the industry that enforcing RES such as wind and solar generation providing PFC with narrow deadband (e.g., $\pm$15 mHz) is contributing to new phenomena arising in power systems (so far unexplained) such as low-frequency oscillations and asymmetry in the frequency distribution \cite{aemopfc}.  The asymmetry of FPD has, for example, a direct impact on devices and/or industrial processes that rely on the use of synchronous electric clocks for time keeping as frequency will tend to spend more time below or above the nominal.  Specifically, the Australian Energy Market Operator (AEMO) has recently acknowledged feedback from industry that: ``\textit{the universal application of very narrow governor deadbands may be contributing to unexplained oscillations from some plant and asymmetry in the NEM’s frequency characteristic.''}, and considers this an issue worth exploring further \cite{aemopfc}.  In addition, the Australian Energy Council (AEC) and, in general, power system industry in Australia, is concerned that this asymmetry is decreasing power system resilience as reported in \cite{csenergy}: ``\textit{Furthermore, since the introduction of mandatory PFR, power system frequency has been exhibiting behaviour that suggests resilience has decreased. As confirmed in
the report prepared by Provecta commissioned by the Australian Energy Council
(AEC) as part of its submission to the Draft Determination, system-wide frequency
is displaying:
\begin{itemize}
\item A ``wobble'' in terms of a slow frequency cycling with a period 18-24 seconds.
\item A ``skew'' in terms of an asymmetry in the distribution.
\end{itemize}
Thus, it is difficult to see how any contribution to improved power system resilience is realised particularly in view of the above comments.}''

These excerpts indicate that the changing dynamic behavior of power systems (which can manifest as the appearance of new phenomena such as asymmetry of the FPD) is yet to be fully understood \cite{8450880}.  References \cite{wen2023non, schafer2018non} are among the few recent works that have studied the issue of asymmetry of FPD in power systems.  Both works utilize the skewness parameter $\beta_{f}$ to measure asymmetry of the distribution.  However, these works aim at identifying the issue rather than fully explain its causes.  Moreover, these works do not propose ways to compensate the asymmetry.  This paper aims at filling these gaps.

This work contributes towards the understanding of the causes of the asymmetry of FPD and proposes how to measure and compensate it.  Specific novel contributions are as follows.
\begin{itemize}
\item Provide analytical insights into the nonlinear relationship between wind speed of wind turbines and system losses and frequency deviation.
\item Propose a metric to quantify the level of asymmetry in power systems.  This metric is the difference between left and right standard deviations of the FPD.  It allows evaluating the ``quality'' of frequency control of different power systems without the need of complex techniques/methods (i.e., only frequency measurements are required that are widely available).
\item Study the sources of asymmetry in power systems, such as losses, control  limits, and wind generation providing PFC and Active Power Control (APC).  The latter is a PFC with a tight ($\pm$ 15 mHz) deadband \cite{eirgrid1}.
\item Show that Automatic Generation Control (AGC) is beneficial to reduce FPD asymmetry.
\item Propose a new nonlinear compensation for PFC and show its effectiveness to reduce the FPD asymmetry.
\item Show through real-world data obtained from the Irish and Australian transmission grids and dynamic stochastic simulations that RES (in particular, wind generation) are a source of asymmetry.
\end{itemize}

The remainder of the paper is organized as follows.  Section \ref{sec:theory} provides a qualitative appraisal of the causes of the asymmetry in the frequency distribution of power systems and presents the proposed control strategy to compensate this asymmetry.  Section \ref{sec:metric} defines various metrics, including the one proposed in this paper, to evaluate the asymmetry of the frequency distribution.  Section \ref{sec:real} illustrates two real-world examples, namely the Irish and Australian transmission systems, where the FPD asymmetry has been observed as a direct consequence of the high penetration of wind power generation.  Section \ref{sec:case} describes several scenarios based on the WSCC 9-bus system benchmark system and illustrates the proposed metrics and compensation control.  Finally, Section \ref{sec:conclu} draws conclusions and outlines future work.

\section{Qualitative Appraisal of Asymmetry}
\label{sec:theory}

In this section, first we show that the cause of the skewness of the distribution of the states of a power system is due to the nonlinearity of the equations of the system itself (Section \ref{sub:covariance}).  This section shows that only for linear systems a Gaussian noise source leads to Gaussian distributions of the system variables.  This conclusion applies to systems of any size.  Then, we discuss two specific examples of nonlinearity that lead to the asymmetry of the FPD, namely network losses (Section \ref{sub:losses}) and wind frequency control obtained with variable pitch angle blades (Section \ref{sub:wind}).  As the source of the asymmetry is the nonlinearity, considering a simple system for the theoretical appraisal helps understanding the problem without lack of generality.  We also briefly discuss the effect of regulator hard limits (Section \ref{sub:limit}) as well as introduce the proposed control-based nonlinear compensation to reduce FPD asymmetry (Section \ref{sub:comp}).

\subsection{Probability Distribution of System Variables}
\label{sub:covariance}

The power system model considered in this work is the following set of Stochastic Differential Algebraic Equations (SDAEs) \cite{6547228}:
\begin{align}
  \Dt{\bfg x} &= \bfg f(\bfg x, \bfg y, \bfg \eta) \, , \label{eq:xeq} \\
  \bfg 0_m &= \bfg g(\bfg x, \bfg y, \bfg \eta) \, , \label{eq:yeq} \\
  \Dt{\bfg \eta} &= \bfg a(\bfg \eta) + \mathcal{B}(\bfg \eta) \, \bfg \xi \, .\label{eq:neq}
\end{align}
The transient behavior of electric power systems is traditionally described through the set of DAEs in \eqref{eq:xeq}-\eqref{eq:yeq}.  $\bfg f$ ($\bfg f: \mathbb{R}^{n+m+p} \mapsto \mathbb{R}^n$) are deterministic differential equations; $\bfg g$ ($\bfg g: \mathbb{R}^{n+m+p} \mapsto \mathbb{R}^m$) are the algebraic equations; $\bfg x$ ($\bfg x\in \mathbb{R}^n$) are the deterministic state variables and $\bfg y$ ($\bfg y \in \mathbb{R}^m$) are the algebraic variables.  The stochastic processes $\bfg \eta$ ($\bfg \eta \in \mathbb{R}^p$) are expressed as set of stochastic differential equations, where $\bfg \xi$ ($\bfg \xi \in \mathbb{R}^q$) is the vector of white noise processes that represent the time derivatives of the Wiener processes.  The stochastic processes are defined by a drift $\bfg a$ ($\bfg a: \mathbb{R}^{n+m+p} \mapsto \mathbb{R}^p$) and a diffusion term $\mathcal{B}$, where $\mathcal{B}$ is a $p \times q$ matrix. 
In the remainder of this work (except for the last scenario of the case study), we assume that $\bfg \eta$ are normally distributed and represent variations of load power consumption or of the wind speed.   

It is possible to show that if the system is linear and $\bfg \eta$ are normally distributed, then also $\bfg x$ and $\bfg y$ are normally distributed.  In fact, a linear system is written as:
\begin{equation}
  \label{eq:linear}
  \begin{aligned}
  \Dt{\bfg x} &= \bfb F_{xx} \bfg x + \bfb F_{xy} \bfg y + \bfb F_{x\eta} \bfg \eta \, , \\
  \bfg 0_m &= \bfb G_{yx} \bfg x + \bfb G_{yy} \bfg y + \bfb G_{y\eta} \bfg \eta \, , \\
  \Dt{\bfg \eta} &= \bfb A_{\eta \eta} \bfg \eta + \bfb B_{\eta \xi} \, \bfg \xi \, ,
  \end{aligned}
\end{equation}
where all matrices are time-invariant.  This system can be rewritten as:
\begin{align}
  \nonumber
  \begin{bmatrix}
    \Dt{\bfg x} \\
    \Dt{\bfg \eta}
  \end{bmatrix}
  &=
    \begin{bmatrix}
      \bfb F_{xx} - \bfb F_{xy} \bfb E  & 
      \bfb F_{x\eta} - \bfb F_{xy} \bfb E \\
      \bfg 0_{p,n} & \bfg A_{\eta \eta}\\
    \end{bmatrix}
  \begin{bmatrix}
    \bfg x \\
    \bfg \eta
  \end{bmatrix}    
  +
  \begin{bmatrix}
    \bfg 0_{n,q} \\
    \bfb B_{\eta \xi}
  \end{bmatrix} \bfg \xi\\
  &=\bfb A \, \bfg z + \bfb B \, \bfg \xi \, ,
  \label{eq:SDAElinear1}
\end{align}  
where $\bfb E = \bfb G^{-1}_{yy} \bfb G_{yx}$ and $\bfg z = [\bfg x^T, \bfg \eta^T]^T$ and the algebraic variables $\bfg y$ have been eliminated using the expression:
\begin{equation}
   \label{eq:algeb}
   \bfg y = 
   - \begin{bmatrix} 
       \bfb G^{-1}_{yy} \bfb G_{yx} &
       \bfb G^{-1}_{yy} \bfb G_{y\eta}
       \end{bmatrix} \bfg z \, .
\end{equation}
Then, the Fokker-Planck equation allows writing the following probability distribution of all state variables in stationary condition \cite{carmichael2013statistical}:
\begin{equation}
  \bfb P(\bfg z) = ({\rm det}\mid 2\pi\bfb C\mid )^{-1/2}\cdot
  {\rm exp}\bigg(-\frac{1}{2}\bfg z^T \bfb C^{\bfg -1} \bfg z\bigg) \, ,\label{eq:pdf}
\end{equation}
where $\bfb C$ is the co-variance matrix of the state variables in \eqref{eq:SDAElinear1} obtained as the solution of the following Lyapunov equation:
\begin{equation}
  \bfb A \bfb C + \bfb C \bfb A^T = - \bfb B \bfb B^T \, . \label{eq:lyapunov}
\end{equation}
The co-variance matrix is symmetrical and positive and it can be interpreted as a linear operator that, if applied to a vector of random variables, maps the linear combination of these random variables.  This property allows to state that the distribution of the states $\bfg x$ can be obtained (for a linear system) directly from the distribution of $\bfg \eta$, as follows:  
\begin{equation}
  \bfg x = \bfb C_{x \eta}
  \bfb C_{\eta \eta}^{-1} \, \bfg \eta \, ,
\end{equation}
where we have clustered matrix $\bfb C$ as: 
\begin{equation}
  \bfb C = \begin{bmatrix}
    \bfb C_{x x} & \bfb C_{x \eta}\\
    \bfb C_{x\eta}^T & \bfb C_{\eta \eta}
  \end{bmatrix} , \label{eq:c}
\end{equation}
where $\bfb C_{x x}$ and $\bfb C_{\eta \eta}$ are the co-variance matrices of $\bfg x$ and $\bfg \eta$, respectively, and $\bfb C_{x \eta}$ represents the co-variance matrix between deterministic $\bfg x$ and stochastic $\bfg \eta$ state variables. 

In conclusion, for a linear system, $\bfg x$ are normally distributed if $\bfg \eta$ are normally distributed.  And, from \eqref{eq:algeb}, one obtains that also $\bfg y$ are normally distributed.
However, the power system model \eqref{eq:xeq}-\eqref{eq:neq} is not linear.  Thus, it has to be expected that nonlinearity \textit{deforms} the distribution of states and algebraic variables.  If the nonlinearity is not symmetrical, it has also to be expected that the distribution of (at least some) variables is skewed.  The following two sections demonstrate this statement through two relevant cases, which are also further illustrated in the case study.

\subsection{Asymmetry due to Network Losses}
\label{sub:losses}

\begin{figure}[htb]
  \begin{center}
    \resizebox{0.7\linewidth}{!}{\includegraphics{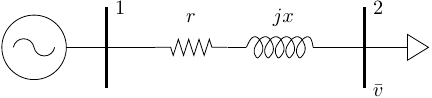}}
    \caption{One-machine one-load system.}
    \label{fig:simple}
  \end{center}
\end{figure}

Let us consider a simple one-machine one-load power system as shown in Fig.~\ref{fig:simple}. Let us also assume that the system is at an equilibrium point.  The power generated by the machine compensates both the load consumption and the losses in the transmission system:
\begin{equation}
   p_G = p_L + p_{\rm loss} \, ,
\end{equation}
where losses are proportional to the square of the current.  Assuming negligible the variations of the voltage, we can write:
\begin{equation}
   \label{eq:pg0}
   p_G \approx p_L + \frac{r}{v^2} \, p_L^2 \, ,
\end{equation}
where $r$ is resistance of the line and the internal losses of the generator and $v$ is the voltage, which, for simplicity of illustration, we assume constant.  Moreover, we assume that initially the frequency of the system is equal to the synchronous reference, namely, $\omega = 1$ pu.

Starting from the initial balance in \eqref{eq:pg0}, consider a variation of the load consumption $\Delta p_L$:
\begin{equation}
   \label{eq:pg1}
   \begin{aligned}
   p_G + \Delta p_G &= p_L + \Delta p_L + \frac{r}{v^2} \, (p_L + \Delta p_L)^2 \, , \\
   &= p_L + \frac{r}{v^2} \, p^2_L + 2\frac{r}{v^2} \, p_L \Delta p_L + \frac{r}{v^2} \, \Delta p_L^2 \, ,
   \end{aligned}
\end{equation}
where $\Delta p_G$ is the active power variation of the generator due to the turbine governor.  
Assuming that $\Delta p_G$ is a function of the droop-control of the turbine governor of the machine, in steady-state, one has:
\begin{equation}
  \label{eq:pg3}
  \Delta p_G = -\frac{1}{R} \, \Delta \omega \, ,
\end{equation}
where $R$ is the droop coefficient of the turbine governor.  For simplicity, we have neglected the contribution of secondary frequency regulation (e.g., AGC) in \eqref{eq:pg3}.   The case study section shows that AGC helps reducing asymmetry but cannot remove it as it does not remove the nonlinearity of the system.  Merging \eqref{eq:pg0} and \eqref{eq:pg3} in \eqref{eq:pg1}, we obtain:
\begin{equation}
    \label{eq:wg}
    \begin{aligned}
    \Delta \omega &= -\left ( 2\frac{R \, r}{v^2} \, p_L \right ) \, \Delta p_L - \left ( \frac{R \, r}{v^2} \right ) \, \Delta p_L^2 \, , \\
    &= -K_1 \, \Delta p_L - K_2 \, \Delta p_L^2 \, ,
    \end{aligned}
\end{equation}
which shows that the effect of losses, i.e., the quadratic term, on the frequency deviation $\Delta \omega$ is always positive and thus has an asymmetrical effect on the steady-state value of the frequency.  This effect, in turn, impacts the probability distribution of the frequency which is skewed towards the side with $\Delta \omega > 0$ pu, as the effect of losses mitigates negative variations of the load power consumption.

\subsection{Asymmetry due to Wind Frequency Control}
\label{sub:wind}

The mechanical power $p_w$ extracted from the wind is a nonlinear function of the wind speed $v_w$, the rotor speed $\omega_m$ and the pitch angle $\theta_p$.  The mechanical power $p_w$ of the wind turbine can be approximated as: 
\begin{equation} 
  \label{eq:dfig2}
  p_w = \frac{\rho}{2} A_r c_p(\lambda,\theta_p) \, v^3_w \, ,
\end{equation}
where $v_w$ is the wind speed; $\theta_p$ and $\lambda$ are the blade pitch angle and  speed tip ratio, respectively; $A_r$ is the area of the turbine disk; $\rho$ is the air density.  The turbine efficiency function $c_p(\lambda,\theta_p)$ can be approximated as follows \cite{Heier:1998}:
\begin{equation} 
  \label{eq:dfig3}
  c_p = 0.22 \biggl( \frac{116}{\lambda_i}-0.4 \, \theta_p -5 \biggr)
  e^{-\frac{12.5}{\lambda_i}} \, ,
\end{equation}
with
\begin{equation} \label{eq:dfig4}
  \frac{1}{\lambda_i} = \frac{1}{\lambda + 0.08 \, \theta_p} -
  \frac{0.035}{\theta^3_p+1} \, ,
\end{equation}
where numerical coefficients are determined empirically.  There exist several other alternative empirical expressions for $c_p$, e.g., \cite{Slootweg:2003a}. 
%
%
In conventional wind turbines, where the main goal of the generator is to maximize the power extracted from the wind, $\theta_p = 0$ in normal operating conditions and $\theta_p \ne 0$ is utilized exclusively to limit the power of the wind turbine at high wind speeds.

A way to regulate the frequency through wind turbines, equivalent to a power reserve, is to introduce an offset of the pitch angle in normal operating conditions \cite{7380877}.  Then the frequency can be regulated through a droop control:
\begin{equation}
    \dot \theta_p = \frac{1}{R} \Delta \omega - \theta_p \, .
\end{equation}
Considering again the load variation $\Delta p_L$ and assuming for simplicity constant wind speed and thus constant speed tip ratio, and that the power injected into by the wind turbine coincides with the mechanical power $p_w$, that is, neglecting losses and considering an ideal converter control, in steady-state and considering small variations around the equilibrium, one has:
\begin{equation}
  \label{eq:pitchdroop}
  \Delta \theta_p = - \frac{1}{R} \Delta \omega \, ,
\end{equation}
and hence:
\begin{equation}
\label{eq:wind}
\begin{aligned}
   \Delta p_G &= 
   c_1 \, \Delta \theta_p + c_2 \, \Delta \theta^2_p + O( \Delta \theta^3_p) \, , \\
   &= - \frac{c_1}{R} \, \Delta \omega + \frac{c_2}{R^2} \, \Delta \omega^2 + O(\Delta \omega^3) \, ,
\end{aligned}   
\end{equation}
where the coefficient $c_1$ and $c_2$ are obtained from the Taylor series expansion of $c_p$ centered at the initial operating point and $O(\Delta \omega^3)$ is the residual terms of order higher than 2.  Neglecting high-order terms, \eqref{eq:wg} obtained in the previous section can be rewritten as follows:
\begin{equation}
    \label{eq:wg2}
    \begin{aligned}
    c_1 \, \Delta \omega - \frac{c_2}{R} \, \Delta \omega^2 &= -K_1 \, \Delta p_L- K_2 \, \Delta p_L^2 \, ,
    \end{aligned}
\end{equation}
%
%
%
which can be solved for $\Delta \omega$ and is, as expected, a nonlinear expression linking $\Delta \omega$ and $\Delta p_L$.  Note that $c_1 \approx c_2$, and hence $c_1 \ll c_2/R$ as the droop coefficient $R$ is of the order of $10^{-2}$.  Then, \eqref{eq:wg2} can be simplified as:  
\begin{equation}
  \label{eq:wg3}
  \Delta \omega^2 \approx \frac{R \, K_1}{c_2} \, \Delta p_L + \frac{R \, K_2}{c_2} \, \Delta p_L^2 \, .
\end{equation}
%

The examples presented in the case study show that the nonlinearity due to wind frequency regulation obtained using pitch angle control is ``stronger'' than that due to system losses and thus leads to higher asymmetry.

\subsection{Asymmetry due to Regulator Hard Limits}
\label{sub:limit}

Another important source of asymmetry in any controller are the hard limits.  If a controller operates close to one of its limits, in fact, every time that the limit is binding the effect the regulator becomes inactive, thus, leading to a substantially different behavior of the controlled system.  This effect is particularly common in wind generation (with or without PFC) and has been observed specifically in the Irish system when narrow deadbands are enforced in the PFC of wind turbines.  This aspect is further discussed in Sections \ref{sec:irish} and in the case study.

\subsection{Compensation of Asymmetry through Control}
\label{sub:comp}



In this section, we propose a nonlinear deadband function to reduce asymmetry which uses the frequency deviation $\Delta \omega$ as an input signal, as follows:
\begin{equation}
  \label{eq:nonlinear}
  u(\Delta \omega) = (1 - \gamma \Delta \omega)^{\alpha} \Delta \omega^{\beta} \, ,
\end{equation}
where $\gamma$ and $\alpha$ are adjustable parameters.  Substituting the expression \eqref{eq:nonlinear} into \eqref{eq:pg3} one obtains:
\begin{equation}
  \Delta p_G = - \frac{1}{R} u(\Delta \omega) \, ,
\end{equation}
and, hence, \eqref{eq:wg} becomes:
\begin{equation}
  \label{eq:wgcomp}
  (1 - \gamma \Delta \omega)^{\alpha} \Delta \omega^{\beta} = -K_1 \Delta p_L - K_2 \Delta p^2_L \, .
\end{equation}
If one chooses $\alpha = \beta = 1$ and $\gamma = K_2$, then:
\begin{equation}
  \label{eq:wgcomp2}
  \Delta \omega = -K_1 \Delta p_L \, ,
\end{equation}
which leads to a linear (symmetric) controller where nonlinearity due to losses is fully compensated.

\begin{figure}[htb]
  \begin{center}
    \resizebox{0.99\linewidth}{!}{\includegraphics{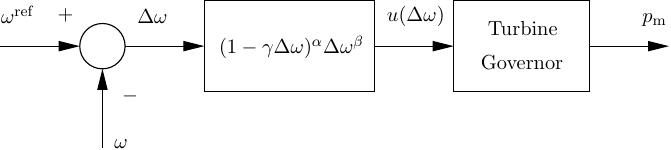}}
    \caption{Proposed nonlinear compensation for turbine governors of synchronous machines.}
    \label{fig:tg}
  \end{center}
\end{figure}

Similarly, if in \eqref{eq:pitchdroop} one implements the proposed controllers, and sets $\alpha = \beta = 0.5$ and $\gamma = R \, K_2/c_2$, \eqref{eq:wg3} can be rewritten as:
\begin{equation}
  \Delta \omega \approx  - \frac{R \, K_1}{c_2} \, \Delta p_L \, ,
\end{equation}
which, again, is a linear expression and makes the pitch-angle frequency control symmetric.

As the expressions \eqref{eq:wg} and \eqref{eq:wg3} are obtained through approximated models where higher order terms are neglected, and as the coefficient $K_2$ depends on the operating point, it is not possible, in general, to obtain a full compensation of the losses.  Based on several tests, for example, we have determined empirically that $\alpha = 0.5$ and $\beta=\gamma=1$ work well in the simulated-based scenario considered in the case study for turbine governors of conventional power plants.


\section{Metrics}
\label{sec:metric}

To measure asymmetry, we propose to calculate the left and right-hand side standard deviations of the FPDs, namely ``negative'' ($\sigma_{f-}$) for when the frequency is below the nominal ($f_n$) (sample size $N_{-}$) and ``positive'' ($\sigma_{f+}$) when frequency is above $f_n$ (sample size $N_{+}$).  Then, we calculate the frequency standard deviation $\sigma_f$ using a weighted-average method of the two standard deviations ($\sigma_{f-}$, $\sigma_{f+}$).  Next, the asymmetry $\Delta \sigma_{f}$ is defined as the difference between $\sigma_{f-}$ and $\sigma_{f+}$.  These equations are shown below:
\begin{align}
  \label{eq:sigma}
  \sigma_{f-} &= \sqrt{ \frac{1}{N_{-}}\sum \limits_{i=1}^{N_{-}} (f_i - f_n)^2 } \, , \; \{ f_i : f_i < f_n \} \, , \\
  \label{eq:sigma1}
  \sigma_{f+} &= \sqrt{ \frac{1}{N_{+}} \sum \limits_{i=1}^{N_{+}} (f_i - f_n)^2 } \, , \; \{ f_i : f_i > f_n \} \, , \\
  \label{eq:sigma2}
  \sigma_{f} &= \sqrt{ \frac{1}{N} \sum \limits_{i=1}^{N} (f_i - f_n)^2 } \approx \sqrt{ \frac{N_{+} \sigma^2_{f+} + N_{-} \sigma^2_{f-} }{N_{+} + N_{-}} } \, , \\
  \label{eq:delta}
  \Delta \sigma_{f} &= \abs{\sigma_{f-} - \sigma_{f+}} \, .
\end{align}
Note that the expressions in \eqref{eq:sigma2} are substantially equal as, in practice, very few measurements are exactly equal to $f_n$.  

As there are other existing asymmetry-based metrics used in the literature such as the skewness parameter, $\beta_{f}$, we calculate and compare it, where relevant, with $\Delta \sigma_{f}$.  The skewness $\beta_f$ is defined as follows \cite{schafer2018non}:
\begin{align}
  \label{eq:beta}
  \beta_{f} &= \frac{1}{N}\sum \limits_{i=1}^{N} \left(\frac{ f_i - \mu_{f}}{\sigma_{f}}\right)^3  \, ,
\end{align}
where $\mu_{f}$ is the mean of system frequency.  Note that if $\beta_{f}$ equals to zero the distribution is Gaussian (symmetric).  In contrast,  a non-zero skewness ($\beta_{f}$) implies a distribution that is not symmetric.

Another key metric of frequency quality used by TSOs is the so-called $\pm$100 mHz criteria which measures the minutes frequency spends outside the $\pm$100 mHz range (i.e., keeping frequency within this range for more than
98\% of the time) \cite{10253411}.  We calculate these minutes in relevant scenarios discussed in Section \ref{sec:case}.

\section{Asymmetry in Real-World Power Systems}
\label{sec:real}

This section provides evidence of FPD asymmetry based on measurements obtained from EirGrid, TSO in Ireland, and from AEMO, TSO in Australia.

\subsection{Real-World Data from the Irish Grid}
\label{sec:irish}
 
We first show the appearance of asymmetry in a real-world system with high shares of wind generation, namely the Irish power system \cite{10253224}.  With this aim, we select three relevant hours with the following details: 
\begin{itemize}
\item Scenario 1: Deadband of wind farms is $\pm$200 mHz and thus APC (i.e., a PFC with a tight $\pm$ 15 mHz deadband) is turned Off.
\item Scenario 2:  Deadband of wind farms is reduced to $\pm$15 mHz and thus APC is turned On.
\item Scenario 3: Represents conventional power systems with near-zero wind generation and APC Off (i.e., $\pm$200 mHz deadband).
\end{itemize}
Further details on each scenario can be found in Table \ref{tab:paramirish}.  Note that currently in the Irish power system APC is normally disabled.  The APC is enabled in special circumstances, e.g., during periods of high export or if there are severe frequency oscillations in the system \cite{eirgrid1}. 

\begin{table*}[t!]
  \centering
  \caption{Scenario description for the Irish power system.}
  \label{tab:paramirish}
  \renewcommand{\arraystretch}{1.2}
  \begin{tabular}{cccccccccccc}
    \hline
    Scenario  & $\rm Wind$ & $\rm APC$ & $\rm fdb_{wind}$ & $\rm fdb_{conv}$  &$\rm AGC$ & $\rm Wind$ & $\rm Load$ & $\rm Wind$ & $\rm Losses$ & $\rm Saturation$ & $\rm fdb_{\rm model}$ \\
     & Generation& & (mHz) & (mHz) & conv/wind & Ramps& Noise& Noise& \\
    \hline
     1  & \OK{} & Off  & $\pm$ 200 & $\pm$ 15 & No & \OK{} & \OK{} & \OK{} & Normal & No & -- \\
    2  & \OK{} & On  & $\pm$ 15 & $\pm$ 15 & No & \OK{} & \OK{} & \OK{} & Normal & No & --\\
    3  & \OK{} & Off  & $\pm$ 200 & $\pm$ 15 & No & \OK{} & \OK{} & \OK{} & Normal & No & -- \\
    \hline
  \end{tabular}
\end{table*}

In the remainder of this section, we consider three measurement data sets obtained from the TSOs historical information system (coming from SCADA and stored in 1 second resolution) and calculate $\sigma_{f}$, $\Delta \sigma_{f}$ and minutes outside the $\pm$100 mHz range.  It is worth mentioning that AGC is not utilized in the Irish power system.  Table \ref{tab:param} summarizes these scenarios.  Further information on the operating conditions for each scenario can be found online in \cite{eirgrid}.  

\begin{table*}[t!]
  \centering
  \caption{Summary of results for the Irish power system.}
  \label{tab:irishresults}
  \renewcommand{\arraystretch}{1.2}
  \begin{tabular}{cccccccccccc}
    \hline
    Scenario  & $\sigma_{f}$ & $\sigma_{f-}$ & $\sigma_{f+}$ & $\Delta \sigma_{f}$  & $\rm Minutes \; Outside$ & $\rm Minutes \; Outside$ & $\rm Minutes \; Outside$ & $\rm P_{loss}$ & $\rm Q_{loss}$ \\
     & (Hz) & (Hz) & (Hz) & (Hz) & $\rm \pm 100mHz$ & $\rm + 100mHz$ & $\rm - 100mHz$ & (pu)& (pu)&\\
    \hline
     1  & 0.0558 & 0.0557  & 0.0560 & 0.0003 & 6.6 & 3.8 & 2.8 & -- & -- \\
    2  & 0.0547 & 0.0259  & 0.0575 & 0.0316 & 7 & 7 & 0 & -- & -- \\
    3  & 0.030 & 0.0359  & 0.0152 & 0.0207 & 0 & 0 & 0 & -- & -- \\
    \hline
  \end{tabular}
\end{table*}

\underline{\textit{Scenarios 1 and 2.}} These two scenarios correspond to the 27 of January 2024, namely 6 consecutive hours, 3 when APC was Off and 3 when APC was turned On.  The Irish power system experienced high wind generation around 3.5 GW.  Note that the peak demand of the Irish system is approximately 7.5 GW (set on 8 January 2025) and valley demand is less than 2.5 GW.  In these 6 hours, the system non-synchronous penetration was around 72\% on average and system conditions remained almost the same, at least, in terms of wind generation, number of conventional units online (i.e., 7) and demand.  
\begin{figure}[thb!]
  \begin{center}
    \resizebox{1.0\linewidth}{!}{\includegraphics{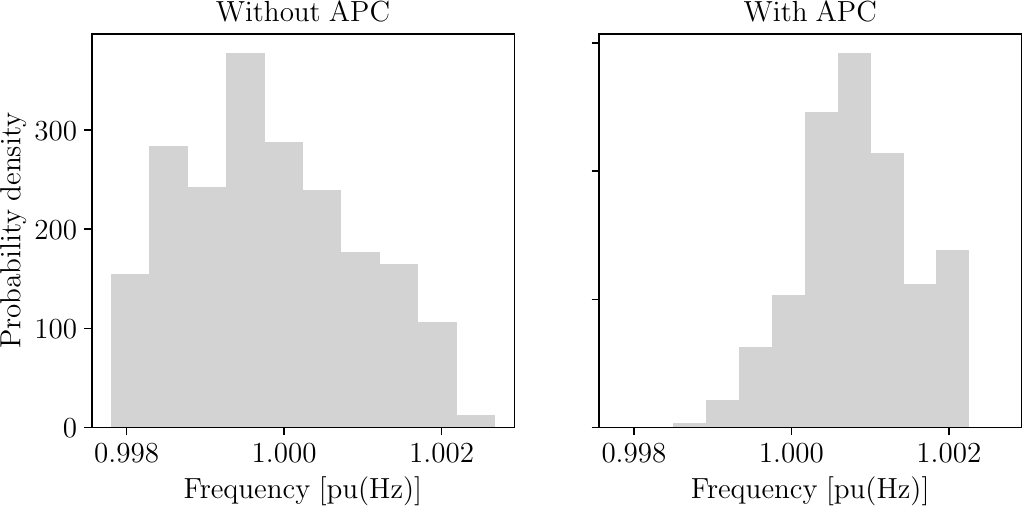}}
    \caption{FPD of the Irish power system with and without APC.}
    \label{fig:irish}
  \end{center}
\end{figure}

Figure~\ref{fig:irish} depicts the  FPDs for both scenarios, namely with and without APC.  APC Off leads to a normal distribution and symmetric behaviour of the FPD.  It can be inferred from Table~\ref{tab:irishresults}, in fact, that the asymmetry for the 3 hours when APC was Off (Scenario 1) is very small and similar to Scenario 4 in the IEEE 9-bus system (see Table~\ref{tab:results} below).  The minutes outside the $\pm$100 mHz range are also small.  It is worth pointing out that whenever the frequency drifts away from the $\pm$100 mHz range, the operators take manual actions (e.g., conventional generation redispatch) to bring back frequency within the range.   

On the other hand, Fig.~\ref{fig:irish} shows that when APC is enabled (Scenario 2) to reduce the frequency oscillations (all 6 APC groups \cite{eirgrid1}) it seriously increases the asymmetry of the FPD (this scenario is equivalent to Scenario 10 in Table~\ref{tab:results} below). 
Specifically, it can be seen from the results in Table \ref{tab:irishresults} that while the average standard deviation ($\sigma_{f}$) is more or less the same with Scenario 1, the asymmetry $\Delta \sigma_{f}$ has increased dramatically.  In fact, the asymmetry is very similar to that considered in Scenario 10 for the IEEE 9-bus system.  The asymmetry can also be observed in the minutes outside the $\pm$100 mHz range.  Note that we tested other real-world scenarios when APC was Off and On and obtained similar results.  Note also that while wind turbines may operate based on nonsymmetric droop characteristic (i.e., which could then be a potential source of asymmetry) \cite{ela2014active}, that is not the case in the Irish power grid (symmetric droop is used instead). 

\underline{\textit{Scenario 3.}} This scenario refers to 20 April 2024 and represents a conventional power system with near-zero wind generation as wind levels in the Irish power system during this particular period were around 50 MW.  This scenario thus allows for an interesting comparison with respect to the Scenarios 1 and 2 with high wind shares (i.e., 3.5 GW). 

Results suggest that  $\sigma_{f}$, $\Delta \sigma_{f}$ and minutes outside the $\pm$100 mHz range are reduced compared to Scenario 2.  However, compared to Scenario 1, the asymmetry ($\Delta \sigma_{f}$) is increased significantly.  This can be explained by the fact that being a near-zero wind generation condition means that the conventional generators online and interconnectors should operate closer to their maximum limits \cite{eirgrid}. 

\subsection{Real-World Data from the Australian Grid}
\label{sec:aus}

As anticipated in the introduction, the asymmetry in the Australian (mainland) power system has recently raised concerns.  We select three years with different deadband implementations namely 2010 ($\pm$30-50 mHz), 2019 ($>$$\pm$150 mHz) and 2023 ($\pm$15 mHz) to compare the asymmetry evolution (see Table~\ref{tab:paramaus}).  Specifically, it is worth pointing out that prior to 2015 the deadband settings in mainland were generally set within a range of $\pm$ 30 mHz to $\pm$ 50 mHz \cite{csenergy}.  Whereas in 2016 the decisions of the Australian Energy Regulator and Market Commission not to enforce generator participants a specific deadband made the latter changing (widening) their deadbands (e.g., $\pm$ 150 mHz), effectively resulting in no PFC in the standard frequency range (i.e., $\pm$ 150 mHz).  Finally, in 2020 the mandatory PFC rule with $\pm$ 15 mHz deadband was enforced motivated by a deterioration of frequency quality.

\begin{table*}[t!]
  \centering
  \caption{Scenario description for the Australian (mainland) power system for years 2010, 2019 and 2023.}
  \label{tab:paramaus}
  \renewcommand{\arraystretch}{1.2}
  \begin{tabular}{cccccccccc}
    \hline
    Year & Wind & APC & $\rm fdb_{wind}$ & $\rm fdb_{conv}$  & AGC & Wind & Wind & Losses & Saturation \\
     & Generation& & (mHz) & (mHz) & conv/wind & Ramps& Noise& \\
    \hline
    2010  & \OK{} & Off  & $\pm$ 30 -- 50 & $\pm$ 30 -- 50 & \OK{} & \OK{} & \OK{} & --  & --\\
    2019  & \OK{} & Off  & $>$ $\pm$ 150 & $>$ $\pm$ 150 & \OK{} & \OK{} & \OK{} & -- & --\\ 
    2023  & \OK{} & On  & $\pm$ 15 & $\pm$ 15 & \OK{} & \OK{} & \OK{} & -- & -- \\
    \hline
  \end{tabular}
\end{table*}

\begin{table*}[t!]
  \centering
  \caption{Summary of asymmetry results for the Australian (mainland) power system for years 2010, 2019 and 2023.}
  \label{tab:australia}
  \renewcommand{\arraystretch}{1.2}
  \begin{tabular}{cccccccccc}
    \hline
    Year  & $\sigma_{f}$ & $\sigma_{f-}$ & $\sigma_{f+}$ & $\Delta \sigma_{f}$  & Minutes Outside & Minutes Outside & Minutes Outside & $\beta_{f}$ \\
     & (Hz) & (Hz) & (Hz) & (Hz) & $\rm \pm 150mHz$ & $\rm + 150mHz$ & $\rm - 150mHz$ & (Hz)&\\
    \hline
    2010  & 0.0286 & 0.0294  & 0.0282 & 0.0011 & 137.13 & 4.4 & 132.73 & -0.18\\
    2019  & 0.0635 & 0.0647  & 0.0643 & 0.00032 & 4059 & 841 & 3218 & -0.047\\
    2023  & 0.0256 & 0.0271  & 0.0244 & 0.00267 & 3.4 & 0.133 & 3.266 & -0.221\\
    \hline
  \end{tabular}
\end{table*}

\begin{figure}[thb!]
  \begin{center}
    \resizebox{1.0\linewidth}{!}{\includegraphics{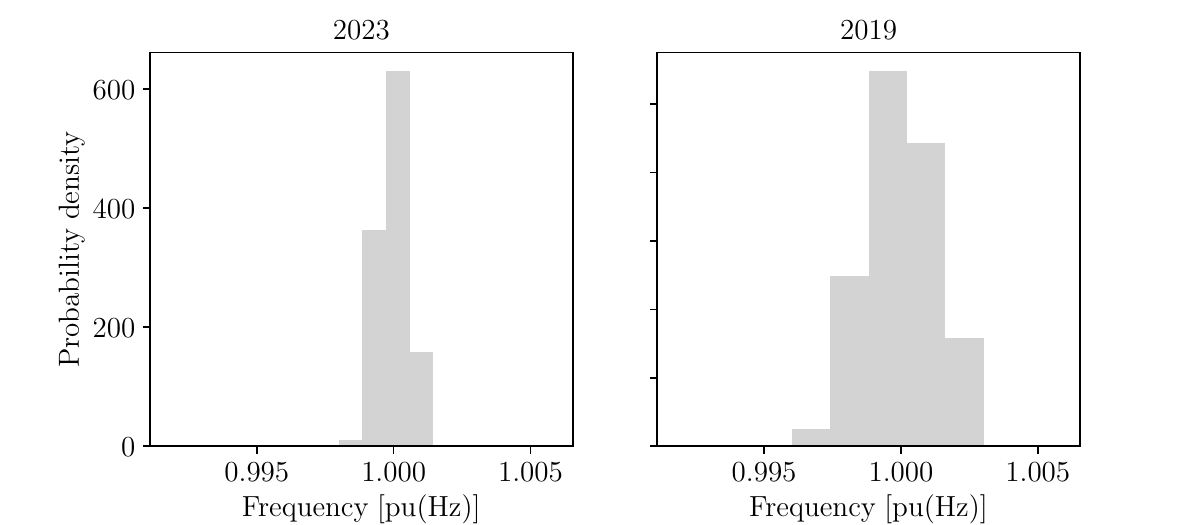}}
    \caption{FPD of the Australian (mainland) power system for 2023 and 2019 years.}
    \label{fig:aus}
  \end{center}
\end{figure}

Table~\ref{tab:australia} presents all the relevant results, while Fig.~\ref{fig:aus} depicts the FPD for years 2023 and 2019 (frequency recordings are made publicly available by AEMO).  It is interesting to observe that while $\sigma_{f}$ and minutes outside $\pm$ 150 mHz have dramatically decreased in 2023 compared to 2010 and 2019, that is not the case for the asymmetry.  Specifically,  the asymmetry for 2023 equals $\Delta \sigma_{f}$= 0.00267 Hz compared to just 0.00032 Hz and 0.0011 Hz in 2019 and 2010, respectively.  As mentioned above, the main change after 2020 is the introduction of mandatory PFC provision with $\pm$ 15 mHz deadband (i.e., APC-type).   The asymmetry in each case can also be observed in the minutes outside $\pm$150 mHz range.  In particular, it is worth mentioning that there is no surprise in the degradation of frequency performance in 2019 (i.e, significant increase in the minutes outside $\pm$150 mHz) given the wider deadband implementation by generators.  These results from the Australian power system further support the theoretical insights and observations in the Irish grid (and also the conclusions of the case study below) that narrow frequency deadbands from RES (e.g., wind generation) lead to substantially increase the asymmetry of the FPD.

Finally, we calculate the skewness parameter $\beta_{f}$ as another metric of asymmetry and to allow comparing it with $\Delta \sigma_{f}$.  Table \ref{tab:australia} suggests that while $\beta_{f}$ is at least two orders of magnitude higher than $\Delta \sigma_{f}$, it is consistent with $\Delta \sigma_{f}$ in terms of it being higher for year 2023 (-0.221 Hz) compared to years 2019 (-0.047 Hz) and 2010 (-0.18 Hz), respectively.  It is also worth pointing out that all $\beta_{f}$ values take a negative sign indicating a leftward skew with a longer tail for negative values of the frequency distribution.  This is also consistent with the results of the standard deviation-based metrics, that is, $\sigma_{f-}$ being higher than $\sigma_{f+}$ in all cases.

\begin{table*}[t!]
  \centering
  \caption{Scenario description for the IEEE 9-bus system.}
  \label{tab:param}
  \renewcommand{\arraystretch}{1.2}
  \begin{tabular}{ccccccccccc}
    \hline
    Scenario & Wind & APC & $\rm fdb_{wind}$ & $\rm fdb_{conv}$  & AGC & Wind & Wind & Losses & Saturation & Compensation \\
     & Generation & & (mHz) & (mHz) & conv./wind & Ramps & Noise & & & Eq.~\eqref{eq:nonlinear} \\
    \hline
    4  & \NO{} & --  & -- & $\pm15$ & conv. & \NO{}  & -- & Normal & \NO{} & \NO{} \\
    5  & \NO{} & --  & -- & $\pm15$ & conv. & \NO{} & -- & High & \NO{} & \NO{}\\ 
       
    6  & \NO{} & --  & -- & $\rm \pm15$ & conv. & \NO{} & -- & High &  \NO{} & \OK{}  \\
    7  & \NO{} & --  & -- & -- & \NO{} & \NO{} & -- & Normal & \OK{} & \NO{}\\
    8  & \OK{} & Off  & $\pm200$ & -- & \NO{} & \NO{} & Gaussian & Normal & \NO{} & \NO{}\\
    9  & \OK{} & On  & $\pm15$ & -- & \NO{} & \NO{} & Gaussian & Normal & \NO{} & \NO{}\\
    10  & \OK{} & On  & $\pm15$ & -- & \NO{} & \OK{} & Gaussian & Normal & \NO{} & \NO{}\\
    11  & \OK{} & On  & $\pm15$ & -- & conv. & \OK{} & Gaussian & Normal & \NO{} & \NO{}\\
    12  & \OK{} & Off  & $\pm200$ & -- & conv. \& wind & \OK{} & Gaussian & Normal & \NO{} & \NO{}\\
    
    13  & \OK{} & On  & $\pm15$ & -- & conv. \& wind & \OK{} & Gaussian & Normal & \NO{} & \NO{}\\
    
    14  & \OK{} & On  & $\pm15$ & -- & conv. \& wind & \OK{} & Gaussian & Normal & \NO{} & \OK{} \\
    
    15  & \OK{} & On  & $\pm15$ & -- & conv. \& wind & \OK{} & Weibull & Normal & \NO{} & \NO{}\\
    \hline
  \end{tabular}
\end{table*}

\begin{table*}[t!]
  \centering
  \caption{Summary of simulation results for the IEEE 9-bus system.}
  \label{tab:results}
  \renewcommand{\arraystretch}{1.2}
  \begin{tabular}{ccccccccccc}
    \hline
    Scenario  & $\sigma_{f}$ & $\sigma_{f-}$ & $\sigma_{f+}$ & $\Delta \sigma_{f}$  & Minutes Outside & Minutes Outside & Minutes Outside & $\rm P_{loss}$ & $\rm Q_{loss}$ \\
     & (Hz) & (Hz) & (Hz) & (Hz) & $\rm \pm 100mHz$ & $\rm + 100mHz$ & $\rm - 100mHz$ & (pu)& (pu)&\\
    \hline
    4  & 0.0107 & 0.0108  & 0.0107 & 0.0001 & 0 & 0 & 0 & 0.0409 & -0.9452  \\
    5  & 0.0314 & 0.0330  & 0.0294 & 0.0036 & 5.28 & 1.27 & 4.00 & 0.5097 & -0.6151  \\ 
    
    6  & 0.0350 & 0.0349 & 0.0350 & 0.00007 & 10.8 & 6.44 & 4.35 & 0.5097 & -0.6151 \\
    7  & 0.0236 & 0.0245 & 0.0227 & 0.0018 & 0.2233 & 0.0183 & 0.205 & 0.0409 & -0.9452 \\
    8  & 0.0602 & 0.0606  & 0.0598 & 0.0008 & 280.49 & 136.38 & 144.10 & -- & -- \\
    9  & 0.0803 & 0.0841  & 0.0762 & 0.0079 & 611.41 & 287.69 & 323.71 & -- & -- \\
    10  & 0.1085 & 0.1254  & 0.0868 & 0.0386 & 1073.61 & 456.67 & 616.94 & -- & -- \\
    11  & 0.0794 & 0.0845  & 0.0745 & 0.01 & 575.79 & 267.32 & 308.47 & -- & -- \\
    12  & 0.0635 & 0.0629  & 0.0641 & 0.0012 & 314.91 & 152.70 & 162.20 & -- & -- \\
    
    13  & 0.0591 & 0.0630  & 0.0555 & 0.0074 & -- & -- & -- & -- & -- \\
    
    14  & 0.0649 & 0.0660  & 0.0638 & 0.0021 & -- & -- & -- & -- & -- \\
    
    15  & 0.1044 & 0.110  & 0.0994 & 0.0105 & -- & -- & -- & -- & -- \\
    \hline
  \end{tabular}
\end{table*}

\section{Case Study}
\label{sec:case}

To analyse the statistical properties of the frequency we run long-term time-domain simulations, namely 48h, based on the SDAE model \eqref{eq:xeq}-\eqref{eq:neq}.  All simulations are solved with the software tool Dome \cite{6672387}.  The results shown in this section are obtained using the IEEE 9-bus system, adapted for the various considered scenarios for a comprehensive evaluation of the different sources of asymmetry.  A short description of each scenario is provided below.  Table~\ref{tab:param} provides relevant information on setup and controllers and Table~\ref{tab:results} shows simulation results for each scenario. 

\begin{itemize}
    \item \underline{\textit{Scenario 4}}: Conventional power systems without wind generation.  Synchronous generators have a $\pm$15 mHz governor deadband and are under AGC. 
    \item \underline{\textit{Scenario 5}}: Same as Scenario 4 but with increase of network losses.
    \item \underline{\textit{Scenario 6}}: Same as Scenario 5 but with inclusion of the proposed nonlinear compensation in the PFC of synchronous generators.
    \item \underline{\textit{Scenario 7}}: Studies the effect of saturation on the FPD.
    \item \underline{\textit{Scenario 8}}: Power system with wind generation providing PFC with $\pm$200 mHz deadband (APC Off).
    \item \underline{\textit{Scenario 9}}: Same as Scenario 8 but with $\pm$15 mHz deadband (APC On).
    \item \underline{\textit{Scenario 10}}: Same as Scenario 9 and with wind ramps modelled based on realistic data from the Irish system.
    \item \underline{\textit{Scenario 11}}: Same as Scenario 10 but with AGC installed (only conventional power plants).
    \item \underline{\textit{Scenario 12}}: Wind farms provide AGC functionality with APC Off.
    \item \underline{\textit{Scenario 13}}: Wind farms provide AGC functionality with APC On.
    \item \underline{\textit{Scenario 14}}: Same as Scenario 13 and APC includes the proposed nonlinear compensation to reduce asymmetry. 
    \item \underline{\textit{Scenario 15}}: Same as Scenario 13 with Weibull-distribution , i.e., asymmetric, wind speed fluctuations.
\end{itemize}

For all scenarios, load consumption includes Gaussian noise modelled as mean-reverted Ornstein-Uhlenbeck processes and stochastic jumps as described in \cite{6547228} and \cite{8827660}, respectively.

\subsection{Power Systems with Conventional Generation}
\label{sec:conv}

In this first section, we focus on the asymmetry that exists in conventional power systems.  Two potential sources of asymmetry are considered, namely, losses and control limiters. 

\begin{figure}[t!]
  \begin{center}
    \resizebox{1.0\linewidth}{!}{\includegraphics{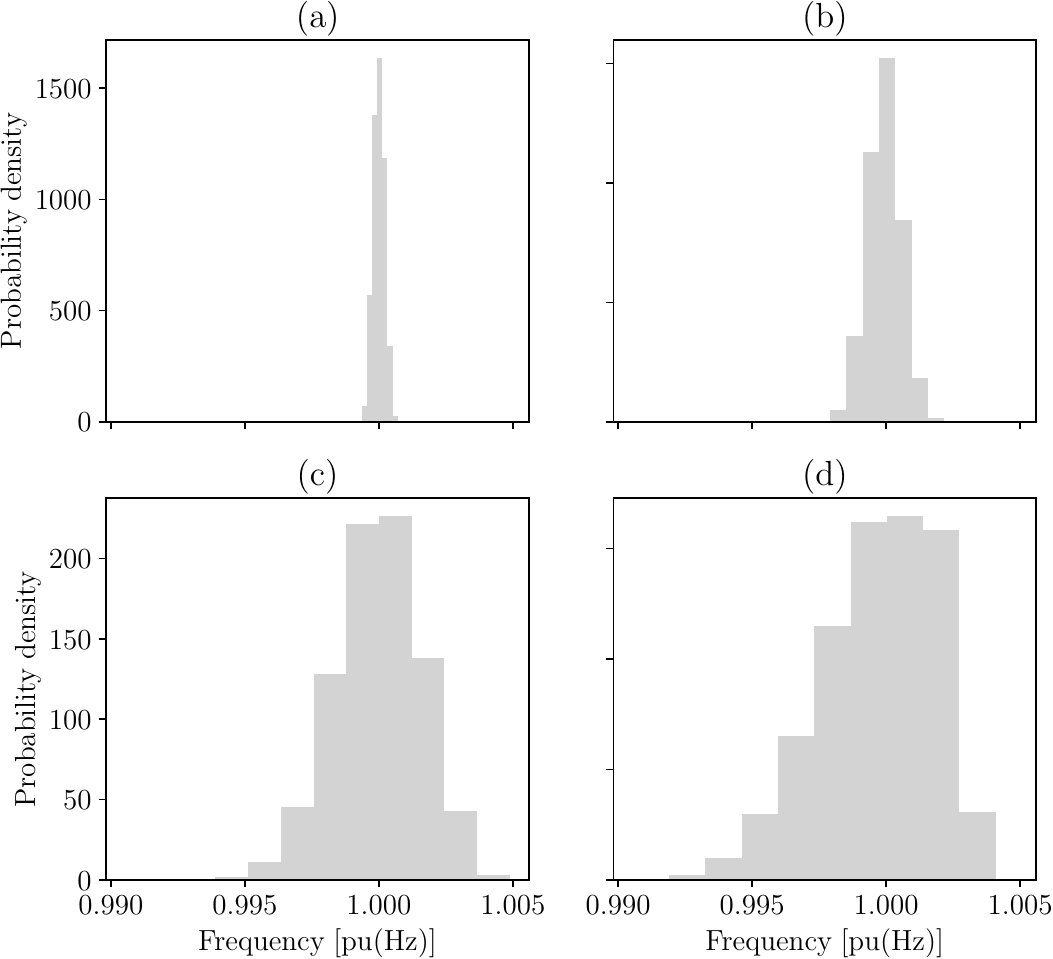}}
    \caption{PD of: (a) Scenario 4, conventional power systems; (b) Scenario 5, high losses; (c) Scenario 9, APC On and no wind ramps; and (d) Scenario 10, APC On and wind ramps.}
    \label{fig:conv1}
  \end{center}
\end{figure}

\underline{\textit{Scenario 4}}: This scenario represents conventional power systems with synchronous machines providing both PFC and AGC \cite{9361269}.  The main source of volatility in this scenario is noise in loads modelled as a stochastic process that incorporates both continuous and event-driven (jumps) dynamics.  This is based on real-world behaviour of loads in the Irish power system \cite{8827660}.  Figure~\ref{fig:conv1}(a) depicts the center of inertia FPD.  As expected, the FPD shows a normal/Gaussian distribution and symmetric behavior.  These results are confirmed in Table \ref{tab:results}, which shows that the asymmetry (e.g., due to network losses) calculated using \eqref{eq:delta} is small ($\Delta \sigma_{f} = 0.0001$ Hz) when compared to power systems with non-synchronous generation (see below).  The minutes outside the $\pm$100 mHz range are zero.  Note that since the focus of this paper is on long-term frequency distribution, it does not matter what frequency measurement is used, that is, the center of inertia or, for example, the frequency measured at a local bus.  In fact, in the Irish grid example, we utilize the frequency measurements at a specific bus.  However, the asymmetry is still present and consistent when using the center of inertia method in this section.

\begin{figure}[h!]
  \begin{center}
    \resizebox{0.99\linewidth}{!}{\includegraphics{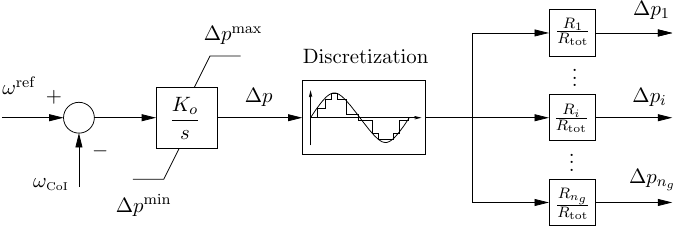}}
    \caption{AGC control diagram.  The discretized output signals $\Delta p_i$ are sent to the PFC of synchronous generators and wind power plants.}
    \label{fig:agc}
  \end{center}
\end{figure}

With regard to AGC, it is worth mentioning that its main goal is to eliminate the frequency error, that is, difference between the reference frequency and the measured frequency at a pilot bus of the system, through an integral controller as shown in Fig.~\ref{fig:agc} \cite{9361269}.  As it shrinks the frequency distribution (see, e.g., \cite{7540970}), the AGC is beneficial to reduce the effects of the asymmetries present in the system.

\underline{\textit{Scenario 5}}: Here we focus on the impact that the increase of system losses have on the FPD.  For illustration purposes, we increase the resistance of transmission lines by ten times (this can represent distribution systems).  Figure~\ref{fig:conv1}(b) shows that losses have a significant impact on the FPD and its asymmetry.  Moreover, Table~\ref{tab:results} shows that, compared to Scenario 4, $\Delta \sigma_{f}$ has increased from $\Delta \sigma_{f} = 0.0001$ Hz to $\Delta \sigma_{f} = 0.0036$ Hz.  Consequently, the standard deviation and the $\pm$100 mHz criteria have also increased.

\underline{\textit{Scenario 6}}: This scenario explores the effectiveness of the proposed nonlinear compensation model presented in Section~\ref{sub:comp} in removing/reducing the asymmetry.  With this aim, we select the case with high losses (Scenario 5) and set the values of $\gamma$ and $\alpha$ to 1 and 0.5, respectively (of course, other values might be used).  Results from Table~\ref{tab:results} indicate that the asymmetry is reduced significantly (from $\Delta \sigma_{f}$ = 0.0036 Hz to $\Delta \sigma_{f}$ = 0.00007 Hz).  Therefore, this compensation can be considered as a viable option for TSOs operating grids with high asymmetry of the frequency distribution.

\underline{\textit{Scenario 7}}: In this scenario, we simulate the effect of saturation/limits (highly loaded systems ).  Specifically, we assume that the maximum powers of synchronous generators 1 and 2 are reduced by approximately 40\% and 50\%, respectively.  As expected, the results of Table \ref{tab:results} indicate an increase in the asymmetry in the FPD compared to, for example, Scenario 4, but lower compared to Scenario 5 that considers high network losses.

\subsection{Power Systems with Non-Synchronous Generation}
\label{sec:modern}

This section discusses the effect of wind generation and its control under different scenarios.  In particular, the aim is to see whether the introduction of wind and its frequency control/regulation capability in the power system affects the asymmetry of the FPD.  With this aim, we consider again the WSCC 9-bus system and replace the synchronous generator 3 with a wind power plant modelled as a double-fed induction generator \cite{Milano:2010}.  We assume that the wind power plant provides primary frequency regulation through droop-based PFC with tight deadband namely $\pm$15 mHz (same used by the governors of conventional generators), as is the case in the Irish power system \cite{10253411}.  This is also known as wind/solar farm APC \cite{eirgrid1}.  To be able to provide up and down regulation, we assume wind is operating 20\% below its maximum power point tracking (i.e., curtailed).  The description of Scenarios 8-15 is provided in Table \ref{tab:param}.

\underline{\textit{Scenario 8}}: We assume that wind farms provide frequency support only to respond to contingency events (deadband of $\pm$200 mHz).  Note that since in practice frequency rarely goes outside the $\pm$200 mHz range during a day (except in very small island power systems and/or microgrids), it can be said that wind farms provide regular wind generation.  We also model noise from wind and loads as Gaussian noise, i.e., $\bfg b$ is a constant vector, but no jumps, i.e., $\bfg c = \bfg 0$, in \eqref{eq:neq}.  Results in Table \ref{tab:results} indicate that the inclusion of wind farms increase $\sigma_{f}$, the asymmetry ($\Delta \sigma_{f} = 0.0008$ Hz) and minutes outside the $\pm$100 mHz range compared to conventional power systems without wind (e.g., Scenario 4).

\underline{\textit{Scenario 9}}: This scenario considers the effect of wind farms providing APC functionality through the $\pm$15 mHz deadband.  This means they are continuously providing dynamic primary frequency regulation since frequency moves outside the $\pm$15 mHz deadband on a regular basis during a day.  Noise is same as in Scenario 8. Figure~\ref{fig:conv1}(c) and Table \ref{tab:results}  show relevant results.  Surprisingly, the APC leads to increase $\sigma_{f}$ and the asymmetry of the FPD (from $\Delta \sigma_{f} = 0.0008$ Hz to $\Delta \sigma_{f} = 0.0079$ Hz).  These results are a consequence of the nonlinearity of the wind turbine, i.e., the cubic relationship between wind speed and turbine torque. 

\underline{\textit{Scenario 10}}: In addition to wind farms providing APC functionality, here we also consider wind ramps modelled based on realistic profile from the Irish power system.  Ramps are obtained through stochastic jumps in \eqref{eq:neq} as described in \cite{8827660}.  Results are shown in Table \ref{tab:results}.  The standard deviation ($\sigma_{f} = 0.1085$ Hz), asymmetry ($\Delta \sigma_{f} = 0.0386$ Hz) and minutes outside the $\pm$100 mHz range (1073.61)  have dramatically increased compared to the previous scenario.  To illustrate this, we plot in Figure~\ref{fig:conv1}(d) the FPD.  It is striking to see that the behavior of the FPD is quite asymmetric. 

\underline{\textit{Scenario 11}}: This is the same scenario as Scenario 10 but now with the inclusion of AGC (only conventional generators).  Looking at the results in Table \ref{tab:results}, we can see now that AGC significantly reduces $\sigma_{f}$ (from 0.1085 Hz to 0.0794 Hz), the asymmetry (from $\Delta \sigma_{f} = 0.0386$ Hz to $\Delta \sigma_{f} = 0.01$ Hz) and minutes outside the $\pm$100 mHz range (from 1073.61 to 575.79).  Thus, the AGC appears as a viable solution to reduce the frequency asymmetry in power systems and help keep frequency quality within limits.

\underline{\textit{Scenario 12}}: The asymmetry of the FPD and dynamic behaviour of the system can be improved if wind is also providing AGC functionality (but with APC Off).  This is shown in Table \ref{tab:results} where we can see that $\sigma_{f}$, $\Delta \sigma_{f}$ and minutes outside the $\pm$100 mHz range are improved significantly with the inclusion of wind farms in AGC.

\underline{\textit{Scenario 13}}: This is the same as Scenario 12 but now we reduce the deadband of wind farms to $\pm$15 mHz (i.e., turn On APC) to see its impact on assymetry.  Interestingly, making wind farms adjust their MW output much more dynamically to control frequency under normal, pre-contingency, conditions leads to an increase of $\Delta \sigma_{f}$ namely from 0.0012 Hz (Scenario 12) to 0.0074 Hz (this Scenario).  These results support the observations in the Irish and Australian power grids where higher asymmetry is observed when RES provide APC-type frequency regulation.

\underline{\textit{Scenario 14}}: In this scenario, we check the effectiveness of the nonlinear deadband implementation in wind farms in reducing the asymmetry.  With this aim, we set the values of $\gamma$ and $\alpha$ to 2 and 0.2, respectively.  Compared to the previous scenario, it can be seen that $\Delta \sigma_{f}$ is reduced by more than three times namely from 0.0074 Hz to 0.0021 Hz.  Therefore, such a deadband implementation might be considered as a potential solution by TSOs.

\underline{\textit{Scenario 15}}: Finally, it is relevant to study the effect of different wind speed noise distributions on the frequency asymmetry, namely, Gaussian (symmetric) and Weibull distribution (asymmetric).  The interested reader can find the details of the model of the wind speed based on Weibull distribution in \cite{Zarate:2013}.  Table~\ref{tab:results} shows that the Weibull noise leads to a higher asymmetry ($\Delta \sigma_{f}$ = 0.0105 Hz) compared to the Gaussian noise ($\Delta \sigma_{f}$ = 0.0074 Hz), as expected.  We note however that, short-term wind speed fluctuations are better represented as Gaussian noise around an average value \cite{JONSDOTTIR2019368}.  Such an average, calculated across periods of, say, an hour, is distributed as a Weibull or other non-symmetrical distributions.  The main source of asymmetry (nonlinearity) due to wind generation is thus not the wind speed but its turbine and its frequency control.

\section{Conclusions}
\label{sec:conclu}

This paper presents a comprehensive analysis of various sources of asymmetry of FPD in power systems.  With this aim, the paper first provides analytical insights on the causes of asymmetrical distribution of the frequency.  Next, the paper proposes a nonlinear compensation suitable for both turbine governors of conventional synchronous machines as well as for the frequency control of variable-speed wind generators. We also propose a new metric based on the difference between the standard deviations of the FPD to evaluate the system asymmetry.  This metric allows consistently comparing asymmetry in different power systems without knowledge of specific parameters and/or system conditions.

Real-world data from the Irish and Australian power systems and a case study based on the IEEE 9-bus system serve to show that RES (e.g., wind generation) are significant sources of asymmetry, in particular, when providing dynamic primary frequency regulation through narrow deadband such as $\pm$ 15 mHz.  \color{black} As mentioned in Section~\ref{sec:intro}, this is already a serious issue being discussed currently in Australia.  The Irish TSOs also recognize the potential appearance of new phenomena as part of the need to have $\pm$ 15 mHz minimum deadband capability from more reserve service providers \cite{dassa}.  It is also shown that AGC is a viable solution to reduce FDP asymmetry.

Future work will focus on evaluating FDP asymmetry of other real-world power systems and explore alternative control-based options to minimize this asymmetry.

\section*{Acknowledgments}

This work was partially supported by Science Foundation Ireland (SFI) by funding F.~Milano under NexSys project, Grant No.~21/SPP/3756;  and by Sustainable Energy Authority of Ireland (SEAI) by funding F.~Milano through FRESLIPS project, Grant No.~RDD/00681.



\vfill

\end{document}